\documentclass[structabstract]{aa}
\usepackage{savesym}
\usepackage{amsmath, bm}
\savesymbol{bibfont}
\usepackage{graphicx}
\usepackage{natbib}
\usepackage{color}
\usepackage{hyperref}
\hypersetup{
     colorlinks=true,
     linkcolor=blue,
     filecolor=blue,
     citecolor = blue,      
     urlcolor=cyan,
     }
\usepackage{txfonts}
\def\be{\begin{equation}}
\def\ee{\end{equation}}
\def\bdm{\begin{displaymath}}
\def\edm{\end{displaymath}}

\begin{document}
	   \title{Numerical simulations of temperature anisotropy instabilities stimulated by suprathermal protons}
	
   \author{S. M.\ Shaaban \inst{1} \fnmsep\thanks{\email{shamd@qu.edu.qa}} R.~A. L\'{o}pez\inst{2,3}  M. Lazar\inst{4,5} \and S. Poedts\inst{5,6}
    }
         \authorrunning{S.\ M.\ Shaaban et al.}
   \institute{
	$^1$ Department of Physics and Materials Sciences, College of Arts and Sciences, Qatar University, 2713 Doha, Qatar\\
	$^2$ Research Center in the intersection of Plasma Physics, Matter, and Complexity ($P^2 mc$), Comisi\'on Chilena de Energ\'{\i}a Nuclear, Casilla 188-D, Santiago, Chile\\
	$^3$ Departamento de Ciencias F\'{\i}sicas, Facultad de Ciencias Exactas, Universidad Andres Bello, Sazi\'e 2212, Santiago 8370136, Chile\\
	$^4$ Institute for Theoretical Physics IV, Faculty for Physics and Astronomy, Ruhr-University Bochum, D-44780 Bochum, Germany\\
	$5$ Centre for Mathematical Plasma Astrophysics, Dept. of Mathematics, KU Leuven, Celestijnenlaan 200B, 3001 Leuven Belgium\\
        $6$ Institute of Physics, University of Maria Curie-Sk{\l}odowska, Pl.\ M.\ Curie-Sk{\l}odowska 5, 20-031 Lublin, Poland
                }
   \date{Received , ; accepted , }

   \abstract
   {The new in situ measurements of the Solar Orbiter mission contribute to the knowledge of the suprathermal populations in the solar wind, especially of ions and protons whose characterization, although still in the early phase, seems to suggest a major involvement in the interaction with plasma wave fluctuations.}
   {Recent studies point to the stimulating effect of suprathermal populations on temperature anisotropy instabilities in the case of electrons already being demonstrated in theory and numerical simulations. 
Here, we investigate anisotropic protons, addressing the electromagnetic ion-cyclotron (EMIC) and the proton firehose (PFH) instabilities.}
   {Suprathermal populations enhance the high-energy tails of the Kappa velocity (or energy) distributions measured in situ, enabling characterization by contrasting to the quasi-thermal population in the low-energy (bi-)Maxwellian core.
We use hybrid simulations to investigate the two instabilities (with ions or protons as particles and electrons as fluid) for various configurations relevant to the solar wind and terrestrial magnetosphere.}
   {The new simulation results confirm the linear theory and its predictions. In the presence of suprathermal protons, the wave fluctuations reach increased energy density levels for both instabilities and cause faster and/or deeper relaxation of temperature anisotropy.
The magnitude of suprathermal effects also depends on each instability's specific (initial) parametric regimes.}
   {These results further strengthen the belief that wave-particle interactions govern space plasmas. These provide valuable clues for understanding their dynamics, particularly the involvement of suprathermal particles behind the quasi-stationary non-equilibrium states reported by in situ observations.}
	

   \maketitle
%

\section{Introduction}\label{sec1}

Plasmas are hot and diluted in many astrophysical contexts, so particle-particle collisions cannot restore thermal equilibrium \citep{Schlickeiser-2002, Marsch2006}. 
Eloquent examples are the solar wind and planetary environments, where in situ measurements reveal non-equilibrium velocity distributions (VDs) of plasma particles (e.g., electrons and protons as major species), which cannot be reproduced by isotropic Maxwellian models specific to thermal equilibrium.
The relevant features are deviations from isotropy (usually relative to the direction of the local magnetic field), such as field-aligned beams and temperature anisotropies, as well as enhanced high-energy or suprathermal tails, modelled by Kappa (or $\kappa$-) power-law distributions \citep{Christon1991, Collier1996, Maksimovic-etal-2005, Stverak2008, Pierrard2010, Anderson-etal-2012, Wang-etal-2015, Lazar-etal-2017, Maruca-etal-2018, Wilson-etal-2019a, Bercic-etal-2019, Durovcova-etal-2021}.

Measured with a cadence lower than the characteristic frequencies of the kinetic (self-generated) instabilities, the anisotropies reveal quasi-stable states, which are, however, sufficiently well shaped by the threshold conditions, suggesting that these instabilities are at work \citep{Kasper2002, Maksimovic-etal-2005, Samsonov-etal-2007, Stverak2008, Bale-2009, Wicks-etal-2013, Lacombe-eta-2014, Maruca-etal-2018, Shaaban-etal-2019b, Shaaban-etal-2019a, Bercic-etal-2019, Huang2020}.
In addition, the suprathermal populations are specific not only to electrons but also to protons and heavier ions \citep{Christon1989, Gloeckler-etal-1995, Collier1996, Lario2019, Yang-etal-2023}, and seem to be the most involved in the mutual interactions with wave fluctuations and plasma instabilities. 
There is also observational evidence for the formation of these suprathermal populations in processes of energization and dispersion by wave fluctuations, self-generated and/or enhanced by kinetic instabilities \citep{Maksimovic-etal-2005, Marsch2006, Bercic-etal-2019, Yang-etal-2023}. With the new missions, Parker Solar Probe and Solar Orbiter, in situ measurements seek to uncover possible coronal sources of suprathermal ions \citep{Bale-etal-2021}, as well as their involvement as seed populations in acceleration processes (and, e.g., generation of energetic particles) as well by waves and plasma turbulence \citep{Wimmer-etal-2021, Mason-etal-2023, Yang-etal-2023}. 

In turn, the suprathermal populations present in the plasma system modify the excitation conditions and the properties of kinetic instabilities, leading, for instance, to the stimulation of those generated by temperature anisotropy \citep{Shaaban-etal-2021-book, Lazar-etal-2022}, or to the inhibition of beam instabilities, in particular, of electrostatic instabilities triggered by electron beams \citep{Lazar-etal-2023}.
Working with non-equilibrium Kappa models is, however, not straightforward.
It requires physically appropriate and observationally consistent approaches to produce quantitative assessments of suprathermal populations and their contributions, from the contrast with the low-energy Maxwellian (core) populations \citep{Lazar2015Destabilizing, Lazar-Fichtner-2021}.
In the case of instabilities generated by temperature anisotropy, that is $A= T_\perp / T_\parallel \ne 1$ (where $\parallel, \perp$ are gyrotropic directions relative to the magnetic field), it is interesting that the suprathermal populations generally have a systematic stimulating effect \citep{Shaaban-etal-2021-book, Lazar-etal-2022}. 
Regardless of the nature of the instabilities, whether electrons or protons generate them, or that they are electromagnetic (EM) cyclotron or mirror modes destabilized by $A>1$, or firehose instabilities induced by $A<1$, linear theory predicts enhanced growth rates and increased ranges of unstable wave-numbers due to suprathermals \citep{Vinas-etal-2015,Lazar2015Destabilizing,Shaaban-etal-2018-mirror,Shaaban-etal-2019a-firehose,Lazar-etal-2019-whistlers,ShaabanPoP2021-emic,ShaabanApJ2021-pfh,Lopez-etal-2023}. 
Implicitly, the anisotropy thresholds of all these instabilities are also reduced.
In the case of electrons, these predictions are also confirmed by theory and nevertheless by numerical simulations, which indicate increases in the wave energy density and a more pronounced effect of fluctuations on the anisotropy relaxation \citep{Lazar-etal-2019-whistlers, Lopez-etal-2019-efh, Moya-etal-2021, Lazar-etal-2022}.

In this paper, we present new results from numerical simulations of protons with anisotropic bi-Kappa distributions in an attempt to confirm theoretical predictions for both the EM ion cyclotron (EMIC) and proton firehose (PFH) instabilities \citep{ShaabanPoP2021-emic, ShaabanApJ2021-pfh}. 
EMIC modes are destabilized by protons (subscript $p$) with $A_p = T_\perp /T_\parallel > 1$, and the parallel (or periodic) PFH instability is triggered by an opposite anisotropy, $A_p < 1$.
Linear and QL properties of these two instabilities have been intensively investigated, especially when triggered by bi-Maxwellian (idealized) protons, such as, for instance, core proton populations in space plasmas, see \cite{Davidson-Ogden-1975, Seough-Yoon-2012, Yoon-Seough-2012, Astfalk-Jenko-2018, ShaabanMNRAS2021} and relevant references in the textbook of \cite{Gary1993}, to name just a few.
Furthermore, numerical simulations can confirm those parametric regimes when these parallel modes are dominant (developing much faster than other instabilities predicted by linear theory), and their effectiveness in relaxing the temperature anisotropy can also be assessed \citep{Gary-etal-1997, Masafumi-etal-2009, Yoon2017, Micera2020, Lopez_2022}. 
Particle-in-cell (PIC) simulations are, however, very demanding in terms of computational resources and have therefore been limited to bi-Maxwellian protons and often unrealistically low proton-to-electron mass ratios, such as $m_p/m_e = 25$, 50, 100 \citep{Gary-etal-2003, Seough2014, Seough2015, Micera2020, Lopez_2022}.
In the present analysis, we use hybrid simulations, which require much less resources and computing time \citep{Gary-etal-1998}.

The results are expected to change markedly under the influence of suprathermal protons modelled with bi-Kappa distributions \citep{Thorne-Summers-1991, Dasso-2003, ShaabanApJ2021-pfh, ShaabanPoP2021-emic, Shaaban-etal-2021-book, Lazar-etal-2022}. 
On the one hand, linear theory predicts a systematic stimulation of both instabilities under the influence of suprathermal protons by enhanced growth rates and lower anisotropy thresholds. 
On the other hand, quasilinear (QL) approaches confirm higher levels of fluctuations reached the saturation and a faster and deeper relaxation of temperature anisotropy \citep{ShaabanApJ2021-pfh, ShaabanPoP2021-emic, Lazar-et al-2022, Lopez-etal-2023}. 
However, numerical simulations still have no definitive confirmation of these effects.
Recent 2D hybrid simulations of EMIC instability assumed bi-Kappa protons to test a new 2D moment-based QL approach, aiming to capture the interplay with the oblique mirror modes \citep{Yoon2023,Lopez-etal-2023}. 
The dominance of EMIC or mirror instabilities depends on the beta parameter, both for bi-Maxwellian \citep{Shaaban-etal-2018-mirror} and bi-Kappa protons \citep{Yoon-etal-2023ApJ}.
Thus, we restrict the analysis to the EMIC dominance conditions (when their growth rates exceed those of the mirror instability), that is, for $\beta_{p\parallel} < 5$, predominant in the solar wind \citep{Wicks-etal-2013} and also relevant for the terrestrial magnetosphere \citep{Maruca-etal-2018}.
The regimes of PFH instabilities are also more favourable to parallel modes \citep{Micera2020, Lopez_2022}, which motivates our restriction to 1D simulations.
This is supported by the observations, which link the parallel-propagating fluctuations at proton scales to instabilities driven by temperature anisotropies and relative ion drifts. In contrast, the oblique fluctuations are consistent with the anisotropic turbulent cascade \citep{Woodham_2019}. 
The simulations of EMIC instability in \citet{Lopez-etal-2023} prove its stimulation by suprathermal protons but only for a limited regime of moderate plasma beta (see definition below), that is $\beta_{p,\parallel} = 1$. 
Here, we survey various regimes of parallel-propagating instabilities of extended relevance in space plasmas, such as the solar wind and planetary environments.

In section~\ref{sec2}, we explicitly introduce VD models used to describe anisotropic protons and isotropic electrons.
The contribution of suprathermal protons to the instability is highlighted by contrasting the results obtained for the bi-Kappa distributed protons with those derived for the bi-Maxwellian limit ($\kappa \to \infty$), reproducing the low-energy core without suprathermal tails \citep{Lazar2015Destabilizing, LazarAA2016}.
Our parameterization relies therefore on the observations featuring the highly dense core, for instance, temperature anisotropy and beta parameter \citep{Kasper2002, Samsonov-etal-2007, Wicks-etal-2013, Maruca-etal-2018, Huang2020}.
The suprathermal protons do not yet benefit from a detailed and systematic characterization (due to limited measurement efficiency and low count statistics). However, ongoing studies provide plausible estimates for the power-law exponent of the Kappa distributions used in our analysis \citep{Nicolaou-etal-2018, Lario2019, Yang-etal-2023}.
Corresponding to the anisotropic models, the linear equations of wave dispersion and stability from kinetic theory are briefly presented in Section~\ref{sec3}.
The unstable solutions of the dispersion relations must describe the instabilities of interest, wave frequencies, and growth rates as functions of wavenumber.
The illustrative results obtained from the simulations for the time evolution of EMIC and PFH instabilities are then discussed in section~\ref{sec4}. 
We analyze the effects of suprathermal protons on the resulting enhanced fluctuations during their growth and saturation. 
Also of interest is the reaction of these fluctuations back on the relaxation of the anisotropic distributions.
In the last section (section~\ref{sec5}), we draw the main conclusions of the present study.

\section{Wave dispersion and stability}

\subsection{Velocity distributions}\label{sec2}

We assume collisionless and homogeneous quasi-neutral electron-proton plasmas, with initially anisotropic protons (subscript $p$), described by the bi-Kappa distribution function \citep{Lazar2015Destabilizing}
\begin{align}\label{eq1}
f_{\kappa,p}\left(v_{\parallel }, v_{\perp }\right)=  \frac{1}{\pi ^{3/2} \theta_{p \parallel} \theta_{p \perp}^{2}} & \,
\frac{\Gamma\left( \kappa +1\right)}{\kappa^{3/2}\Gamma \left( \kappa -1/2\right)}\nonumber\\
&\times \left[ 1+\frac{v_{\parallel }^{2}}{\kappa \theta_{p \parallel }^{2}}
+\frac{v_{\perp }^{2}}{\kappa \theta_{p \perp}^{2}}\right] ^{-\kappa-1}.
\end{align}
This is defined in terms of the (normalizing) velocities $\theta_{p \parallel, \perp}(t)$, which vary with time ($t$) in simulations (see below), and are related to the anisotropic temperature components~(for~any~$\kappa >~3/2$)
\begin{align}
T_{p \parallel}^{\kappa}=&\frac{2\kappa}{2\kappa-3}\frac{m_p \theta^2_{p \parallel}}{2 k_B}>T_{p \parallel} = \frac{m_p \theta^2_{p \parallel}}{2 k_B}, \nonumber \\ 
T_{p \perp}^{\kappa}=&\frac{2\kappa}{2\kappa-3}\frac{m_p \theta^2_{p \perp}}{2 k_B}>T_{p \perp} =\frac{m_p \theta^2_{p \perp}}{2 k_B}, \label{eq2}
\end{align}
as given by the second-order moments of the bi-Kappa distribution. 
Here $m_p$ is the proton mass, and $k_B$ is the Boltzmann constant.
$T_{p \perp, \parallel}$ approximate the temperature components of the lower energy protons in a bi-Maxwellian core (without suprathermal tails) 
\begin{align}\label{eq3}
f_{p}\left(v_{\parallel }, v_{\perp }\right) =&\frac{1}{\pi^{3/2} \theta_{p \parallel}^{2}\theta_{p \perp}}\exp \left(-\frac{v_{\parallel }^{2}}{\theta_{p \parallel}^{2}}
-\frac{v_{\perp }^{2}}{\theta_{p \perp}^{2}}\right),   
\end{align}
which is recovered from \eqref{eq1} in the limit of large $\kappa \rightarrow~\infty$ \citep{Lazar2015Destabilizing}. The results obtained for bi-Kappa distributed protons thus allow a straightforward comparison with those derived in the absence of suprathermal protons in the (bi-)Maxwellian limit, and the difference even characterizes the effects of the suprathermal component \citep{Lazar2015Destabilizing, LazarAA2016}.

In linear theory, we can assume electrons (subscript $e$) to be isotropic and Maxwellian-distributed
\begin{align}\label{eq4}
f_{e}\left( v\right) =&\frac{1}{\pi^{3/2} \theta_{e}^{3}}\exp \left(-\frac{v^{2}} {\theta_e^{2}}\right),   
\end{align}
with thermal velocity $\theta_e$ linked to temperature $T_e = m_e \theta^2_e/(2k_B)$, where $m_e$ is the electron mass.

\subsection{Linear dispersion relations} \label{sec3}
%
\begin{figure*}[t!]
    \centering
    \includegraphics[width=0.9\textwidth, trim={2.cm 7.cm 2.0cm 6.5cm}, clip]{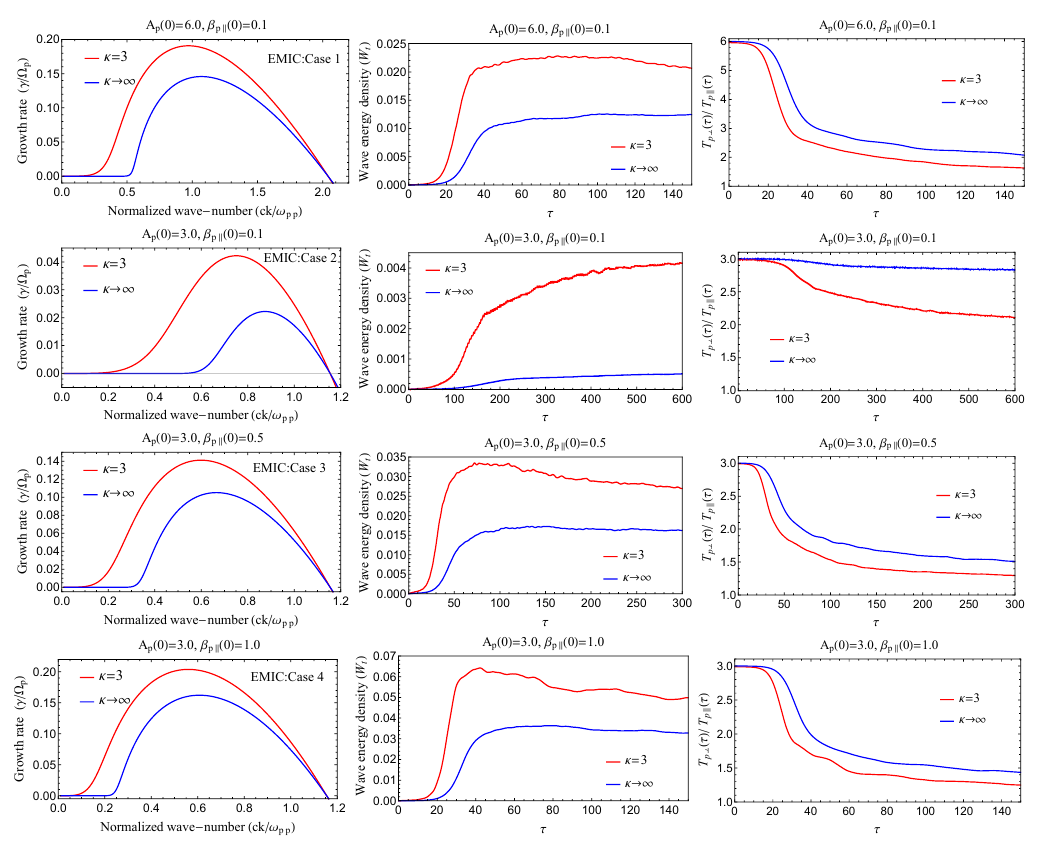}
     \caption{EMIC unstable solutions for bi-Kappa protons (red) and the corresponding bi-Maxwellian core (blue) for all four cases in Table~\ref{t1} with initial proton parameters also indicated in the panels. Comparison of growth rates from linear theory (left panels) and temporal evolution from simulations for the wave energy density (middle panels) and the temperature anisotropy (right panels).}
    \label{f1}
\end{figure*}

\begin{figure}[h!]
    \centering
     \includegraphics[width=0.41\textwidth,clip]{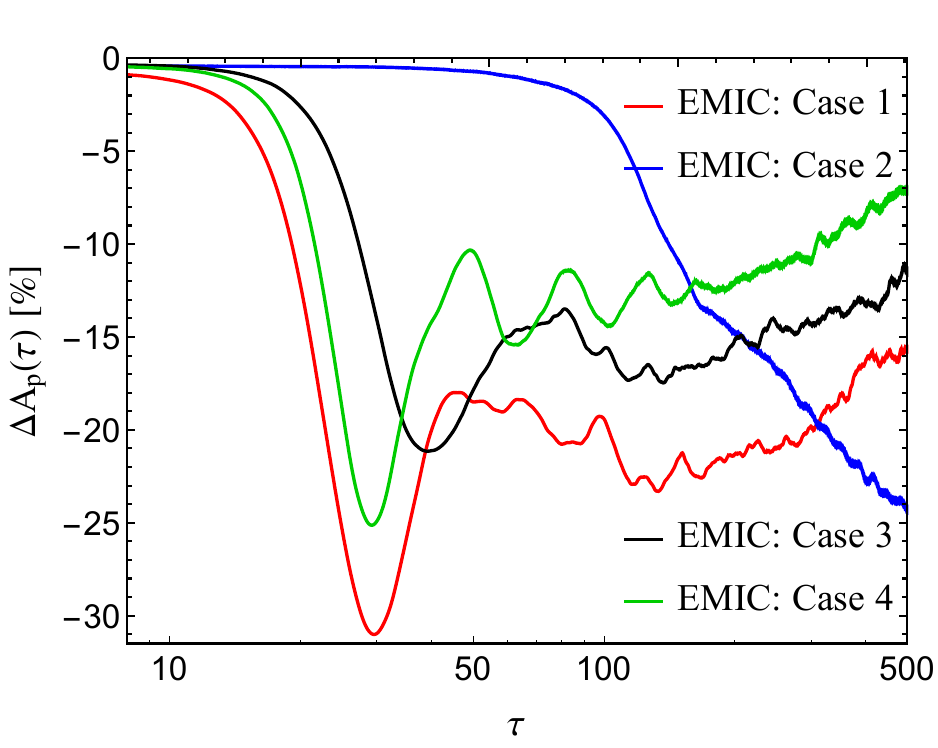}
     \caption{Instantaneous relative difference of the proton anisotropy, $\Delta A_p (\tau)$ (in \%, see definition in the text), induced by the presence of suprathermal protons for all EMIC runs in Fig.~\ref{f1}.}
    \label{f2}
\end{figure}
%

For the plasma configuration described above, the instantaneous dispersion relation for the transverse modes propagating parallel to the background magnetic field reads \citep{ShaabanPoP2021-emic, ShaabanApJ2021-pfh}
\begin{align} \label{eq5}
\tilde{k}^2= &\mu \left[A_e-1+\frac{A_e~\tilde{\omega} \pm \left(A_e-1\right)\mu}{\tilde{k} \sqrt{\mu \beta_{e \parallel}}} Z_{M, e}\left(\frac{\tilde{\omega}\pm \mu}{\tilde{k} \sqrt{\mu\beta_{e \parallel}}}\right)\right] \nonumber\\
&+A_p-1+\frac{A_p~\tilde{\omega} \mp \left(A_p-1\right)}{\tilde{k} \sqrt{\beta_{p \parallel}}}Z_{\kappa,p}\left(\frac{\tilde{\omega}\mp1}{\tilde{k} \sqrt{\beta_{p \parallel}}}\right).
\end{align} 
Here $\tilde{k}=ck/\omega_{p p}$ is the normalized wave-number $k$, $c$ is the light speed, $\omega_{p p}=(4\pi n_p e^2/m_p)^{1/2}$ is the proton plasma frequency (proportional to the number density $n_p=n_0$), $\tilde{\omega}=\omega/\Omega_p$ is the normalized wave frequency $\omega$, $\Omega_j=e B_0/m_j c$ is the non-relativistic gyro-frequency of the plasma species $j$ (with elementary charge $e$), $\beta_{j \parallel}=\theta_{j \parallel}^2 \omega_{pj}^2/(c^2\Omega_j^2) = 8 \pi n_0 k_B T_{j \parallel}/B_0^2$ are the corresponding parallel plasma beta parameters (in the Maxwellian limit, i.e., not depending on $\kappa$), $A_j\equiv \theta^2_{j \perp}/\theta^2_{j \parallel} \equiv T_{j \perp}^\kappa/T_{j \parallel}^\kappa \equiv ~T_{j \perp}/T_{j \parallel}\equiv\beta_{j \perp}/\beta_{j \parallel}$ is the temperature anisotropy of the plasma species $j$, $\mu=~m_p/m_e$ is the proton to electron mass ratio. In the proton term "$\mp$" denotes the circular left-handed (LH) or right-handed (RH) polarization, respectively, and 
\begin{align} \label{eq6}
 Z_{\kappa,p}\left( \xi_{p}^{\mp}\right) =\frac{1}{\pi ^{1/2}\kappa^{1/2}}&\frac{\Gamma \left( \kappa \right) }{\Gamma \left(\kappa -1/2\right) } \nonumber\\  &
 \times \int_{-\infty }^{\infty }\frac{\left(1+x^{2}/\kappa \right) ^{-\kappa}}{x-\xi_{p}^{\mp}}dx,\  \Im \left(\xi _{p}^{\mp}\right) >0.
\end{align}
is the modified dispersion function for Kappa-distributed plasmas \citep{Lazar2008} of argument $\xi_p^{\mp}=~(\omega\mp\Omega_p)/(k \theta_{p \parallel})$. In the Maxwellian limit $Z_{\kappa \to \infty,p}\left( \xi_{p}^{\mp}\right) = Z_{M,p}\left(\xi_{p}^{\mp}\right)$ we recover the standard plasma dispersion function \citep{Fried1961}
\begin{equation}  \label{eq7}
Z_{M,p}\left(\xi_{p}^{\mp}\right) =\frac{1}{\sqrt{\pi}}\int_{-\infty
}^{\infty }\frac{\exp \left(-x^{2}\right) }{x-\xi_{p}^{\mp}}dx,\
\ \Im \left( \xi_{p}^{\mp}\right) >0. 
\end{equation}
In the dispersion relation~\eqref{eq5}, "$\pm$" in the electron term  corresponds to circular LH or RH polarization, respectively (due to the opposite sign of the electric charge), with a similar (standard) dispersion function
\begin{equation}  \label{eq8}
Z_{M,e}\left(\xi_{e}^{\pm}\right) =\frac{1}{\sqrt{\pi}}\int_{-\infty
}^{\infty }\frac{\exp \left(-x^{2}\right) }{x-\xi_{e}^{\pm}}dx,\
\ \Im \left( \xi_{e}^{\pm}\right) >0, 
\end{equation}
of argument $\xi_e^{\pm}=(\omega\pm|\Omega_e|)/(k \theta_{e \parallel})$.

In the absence of suprathermal protons (i.e., $\kappa \to \infty$), the modified dispersion function $Z_{\kappa,p}$ converges to the standard dispersion function $Z_{M,p}$. Consequently, in this limit, the linear dispersion relation \eqref{eq5} reduces to the standard linear equation that describes the instabilities of the (cooler) bi-Maxwellian core \citep{Gary1993, ShaabanMNRAS2021}.
The results from linear theory show the effects of suprathermal protons on the wave frequency and growth rates of EMIC and PFH instabilities. 
For instance, growth rates are displayed for each case of interest in Figs.~\ref{f1} and \ref{f4} (left panels), and wave frequencies in Fig~\ref{f7} to show the agreement of the simulations with predictions of linear theory.

%
\begin{figure*}[h!]
    \centering
    \includegraphics[width=0.85\textwidth,trim={3.5cm 11.5cm 3.5cm 2cm},clip]{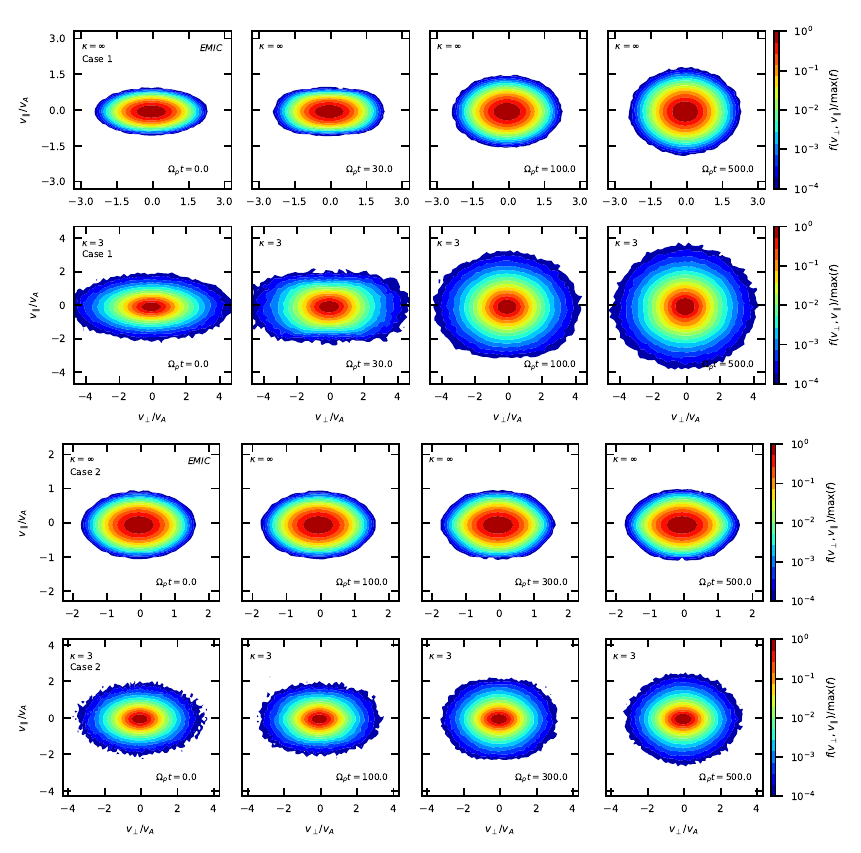}
    \caption{VDs from the simulations of EMIC instability for cases~1 and 2 from Figs.~\ref{f1}, computed for four time steps, and for both bi-Maxwellian ($\kappa \to \infty $) and bi-Kappa ($\kappa=3$) distributed protons, as indicated in the panels.}
    \label{f3}
\end{figure*}
%

\section{Hybrid simulations} \label{sec4}

We have used 1D explicit hybrid simulations to examine the two unstable EM modes: the parallel EMIC and PFH instabilities. In the simulations, protons are modelled as particles with suprathermal, (bi-)Kappa tails and temperature anisotropies $A_p\neq 1$, whereas electrons are considered a massless charge-neutralizing fluid with a constant temperature. This code is based on the \citet{Winske1993} code. Protons are advanced using the Boris scheme~\citep{Boris1970,Boris1970b}, and the fields are computed explicitly following the iterative scheme as in \citet{Horowitz1989}. The initial proton (bi-)Kappa distribution is loaded using the acceptance-rejection method; see details in \citep{Lopez-etal-2023}.
For all investigated cases, we assume isotropic electrons with $\beta_e = 0.5$.
The simulation box contains 4096000 protons within a length $L=502.65\,v_A/\Omega_p$ and $4096$ grid points, and for the normalized time $\tau=\Omega_p t$ a step of $\delta \tau=0.01$ is used. The constant background magnetic field is set in the $z$-direction.

Tables~\ref{t1} and \ref{t2} contain values of the key plasma parameters, $A_p$ and $\beta_{p\parallel}$, used in our simulations. 
These values are inspired by solar wind observations, mainly proton core data reported from various heliocentric distances by various spacecraft \citep{Kasper2002, Wicks-etal-2013, Shaaban-etal-2017, Huang2020}, whose relevance can also extend to the terrestrial magnetosphere \citep{Samsonov-etal-2007, Maruca-etal-2018}.
Moreover, our parameterization is generally favourable to parallel propagating instabilities (see the motivations in section~1).
Observations are quasi-steady states confined to a rhombic-shaped outer boundary known as $(A_p, \beta_{p\parallel})$--parameter space.
Due to a limited measurement cadence, the observed anisotropies are relatively small, specific to quasi-stable states below the thresholds of instabilities.
Larger deviations from isotropy ($A_p=1$) may exist. Still, they cannot be observed (for now) because they are rapidly relaxed by our instabilities, whose growth rates and relaxation effects increase with the initial anisotropy.
This motivates the values we invoke for temperature anisotropies, which can be at or even above the upper limits of the observations.

\subsection{EMIC instability}

First, we discuss the results obtained for the EMIC instability for four distinct cases, explained in Table~\ref{t1} in terms of the plasma parameters' initial ($\tau =0$) values. According to Eqs.~\eqref{eq2}, the (initial) proton anisotropy $A_p$ does not depend on the distribution model, as it happens with the (initial) plasma beta parameters. 
Such that, for initial bi-Kappa distributed protons with $\kappa = 3$ parallel plasma beta becomes two times higher than that for quasi-thermal bi-Maxwellian protons ($\kappa \to \infty$), that is, $\beta_{p\parallel}^{\kappa = 3} (0) = 2 \beta_{p\parallel}(0)$.
Corresponding EMIC solutions are displayed in Fig.~\ref{f1}, with red and blue lines, respectively. 

Left panels in Fig.~\ref{f1} display the normalized growth rates ($\gamma/\Omega_p$) derived from linear theory, which show a systematic enhancement, with higher peaks and larger ranges of unstable wavenumbers in the presence of suprathermal protons (red). These results agree with previous linear approaches of EMIC instability in plasmas with bi-Kappa protons \citep[and refs. therein]{Shaaban-etal-2021-book}.
Corresponding to these two (linear) solutions, panels in the middle column also compare the time evolution of the wave energy densities ($W_t=\delta B^2/B_0^2$), as obtained from simulations. 
The right column shows the time relaxation of the temperature anisotropy, providing the same comparison between the solutions derived for bi-Kappa (red) and bi-Maxwellian (blue) protons. 
%
\begin{table}[t!]
    \caption{Parameterization adopted in the hybrid simulations of the EMIC instability, in terms of initial values ($\tau = 0$) for four distinct runs.} \label{t1}
   \centering 
    \begin{tabular}{@{}lcccc}
    \hline
     Parameters $ \backslash $ Runs & 1 & 2 & 3 & 4\\
    \hline
    $\beta_{p,\parallel}(0)$ & 0.1 & 0.1 & 0.5 & 1.0\\
    $\beta_{p,\parallel}^{\kappa=3} (0) $ & 0.2 & 0.2 & 1.0 & 2.0\\
    $A_p (0)$ & 6 & 3 & 3 & 3\\
    \hline 
    \end{tabular}
\end{table}
\normalsize

First, it should be emphasized that the simulations confirm the linear theory predictions for a wide variety of cases studied (including those not discussed here).
Namely, for higher growth rates of EMIC instability in the presence of suprathermal protons (red lines), with markedly higher peaks of maximum growth rates ($\gamma_m > 0$), the resulting wave fluctuations reach a much higher level of energy density at the saturation (middle panels).
The increase rate in wave energy density is roughly given by  $2\gamma_m$ (see, for instance, Fig.~8). In the presence of suprathermal bi-Kappa protons, it can thus reach saturation values up to several times higher, see cases~2 and 3.
More intense wave fluctuations can also cause and explain the faster and deeper relaxation of the temperature anisotropy shown in the right panels. 
Also, the onset times of the increase in fluctuations and relaxation of the anisotropy can be noticeably shorter.
This is due to the suprathermal protons and the additional energy they contribute, which is partially converted into EMIC waves.

Initial values of the main parameters, namely the plasma beta and temperature anisotropy, affect the evolution of the instability. 
Higher values of the proton beta and anisotropy mean more free energy in the system, leading to higher growth rates of the EMIC instability and larger intervals of unstable wavenumbers. 
This is proven by a direct comparison between runs~1 and 2, which start from the same low $\beta_{p\parallel}(0) = 0.1$, but different anisotropies, respectively, $A_p (0)=6$ and $A_p (0)=3$. 
Higher initial anisotropy in run~1 determines higher levels of fluctuations (quantified by the wave energy density) and deeper relaxation of the temperature anisotropy as time evolves. Similar effects are obtained when the anisotropy is kept the same. Still, the initial plasma beta increases, for instance, if we compare runs 2, 3, and 4, for which we start with the same $A_p (0)=3$, but different $\beta_{p\parallel} (0)=0.1$, 0.5, and 1, respectively. Moreover, we find that all these effects are stimulated by the presence of suprathermal populations, which contribute to the increase of kinetic energy (temperature) of protons and implicitly to an increase of the beta parameter. Thus, growth rates are enhanced, the wave energy density of the resulting EMIC fluctuations is increased, and the relaxation of proton anisotropy is deeper, approaching closer to the isotropy condition.

Up to this point, the impact of suprathermal protons on temperature anisotropy relaxation has not yet been quantified. 
Therefore, in Fig.~\ref{f2}, we examine the instantaneous relative difference of the proton anisotropy (in \%)
\begin{align}
\Delta A_p=\frac{A_p^\kappa(\tau)-A_p^M (\tau)}{A_p^M (\tau)}\times 100, \label{eq9}
\end{align}
as a function of different initial plasma parameters for all cases in Fig.~\ref{f1}. 
$A^\kappa_p(\tau)$ and $A^M_p (\tau)$ represent the instantaneous temperature anisotropies for bi-Kappa and bi-Maxwellian cases. 
In Fig.~\ref{f2}, it is evident that the instantaneous relative difference $\Delta A_p (\tau)$ reaches minima
at the early stages, specifically at $\tau \in [30-40]$, for runs 1, 3, and 4 with highest levels of wave fluctuations. 
These minima indicate that the relaxation of temperature anisotropy starts much earlier due to suprathermal protons and becomes significant even before Maxwellian protons begin to relax. 
Conversely, in run 2, the fluctuations reach only lower levels, and the decrease in $\Delta A_p$ (blue line) starts around $\tau \simeq 80$. 
For run~1 with $A_p(0)=6$ and $\beta_{p\parallel} (0)=0.1$ (red line), $\Delta A_p$ decreases to its minimum value of $-31~\%$ at $\tau \simeq 30$ before eventually reaching a value of $-16~\%$ at $\tau \simeq 500$. 
This means the relaxation of the temperature anisotropy in the presence of suprathermal (Kappa) protons is 31~\% deeper already at $\tau \simeq 30$, and this difference is reduced to 16~\% later at $\tau \simeq 500$.

Runs 3 (black) and 4 (green) with higher initial betas, that is, with $\beta_{p\parallel} (0)= 0.5$ and 1.0, respectively, show similar behaviour as time evolves but with lower differences $\Delta A_p$. For instance, in case~3 $\Delta A_p = -21.5$~\% at $\tau \simeq 40$ and $-11~\%$ at $\tau \simeq 500$, whereas for case~4  $\Delta A_p= -25.2$~\% at $\tau \simeq 29$ and $-7~\%$ at $\tau \simeq 500$. 
One conclusion to draw here is that as proton plasma beta increases, the effects of suprathermal populations on the resulting fluctuations, and implicitly on the temperature anisotropy relaxation. 
This influence also becomes evident even with low levels of wave fluctuations, such as in run~2 (blue line); that is, the instantaneous difference $\Delta A_p=-25$~\% at $\tau_{max}$ is larger than all the other cases at final stages. 
The EMIC instability is typical of magnetized plasmas, so variations of $\beta_{p \parallel}$ have a greater impact on the growth rate and fluctuation levels in highly magnetized plasma (with low $\beta_{p \parallel}$) than in weakly magnetized plasmas (with high $\beta_{p \parallel}$).
This is also reflected in the contribution of suprathermal protons.

Fig.~\ref{f3} displays the simulated proton VDs, namely, snapshots of $f_p(v_\parallel, v_\perp)$, which enable us to identify the deformations that eventually appear compared to the initial shape of these distributions as time evolves, in response to the growth of EMIC fluctuations.
The upper two rows of panels show the bi-Maxwellian ($\kappa\rightarrow \infty$) and bi-Kappa ($\kappa=3$) protons for run~1 in Fig.~\ref{f1} ($A_p(0)=6$ and $\beta_{p\parallel}(0)=0.1$), at four times: $\tau=0, 30, 100, 500$ (including saturation and extending by the end of the simulation).
The initial shape of these distributions, either bi-Kappa or bi-Maxwellian, is generally preserved, although these distributions are subjected to the relaxation of temperature anisotropy.
To capture the differences introduced by the suprathermal protons on the VD evolution, the time step $\tau=30$ is carefully selected in correspondence with the observed minimum in Fig.\ref{f2} (red line).
At $\tau = 30$, one can see that the initial bi-Maxwellian VD undergoes a small deformation, mainly in the sense of relaxation. In contrast, the initial bi-Kappa distribution shows stronger relaxation towards isotropization.
The cyclotron resonant particles are pitch-angle and energy-scattered. As the time progresses to saturation, that is, at $\tau \simeq 100$, the main response of the EMIC fluctuation to the VD is completed.
The anisotropic bi-Kappa relaxes faster than the bi-Maxwellian, showing a less elliptical asymmetry. 
After saturation towards the end of the simulation (i.e., at $\tau=500$), the particle diffusion in velocity space continues, but less prominently, though the VDs in both cases become more isotropic. 
The lower panels in Fig.~\ref{f3} are intended for a similar comparison but for run~2 (blue line in Fig.~\ref{f2}). The contrast between bi-Kappa and bi-Maxwellian distributions becomes evident after $\tau \simeq 100$ increases as time evolves. 
The results for the bi-Maxwellian protons are consistent with those in \cite{Gary-etal-2003}. 
A comparison of the results from runs~1 and 2 reveals that VDs associated with intense wave fluctuations (run~1) relax faster and more efficiently to approach the isotropy than those associated with lower levels of fluctuations (run~2).
%
\subsection{PFH instability}
%
\begin{figure*}[t!]
    \centering
     \includegraphics[width=0.9\textwidth,trim={2.cm 7.cm 2.0cm 6.5cm}, clip]{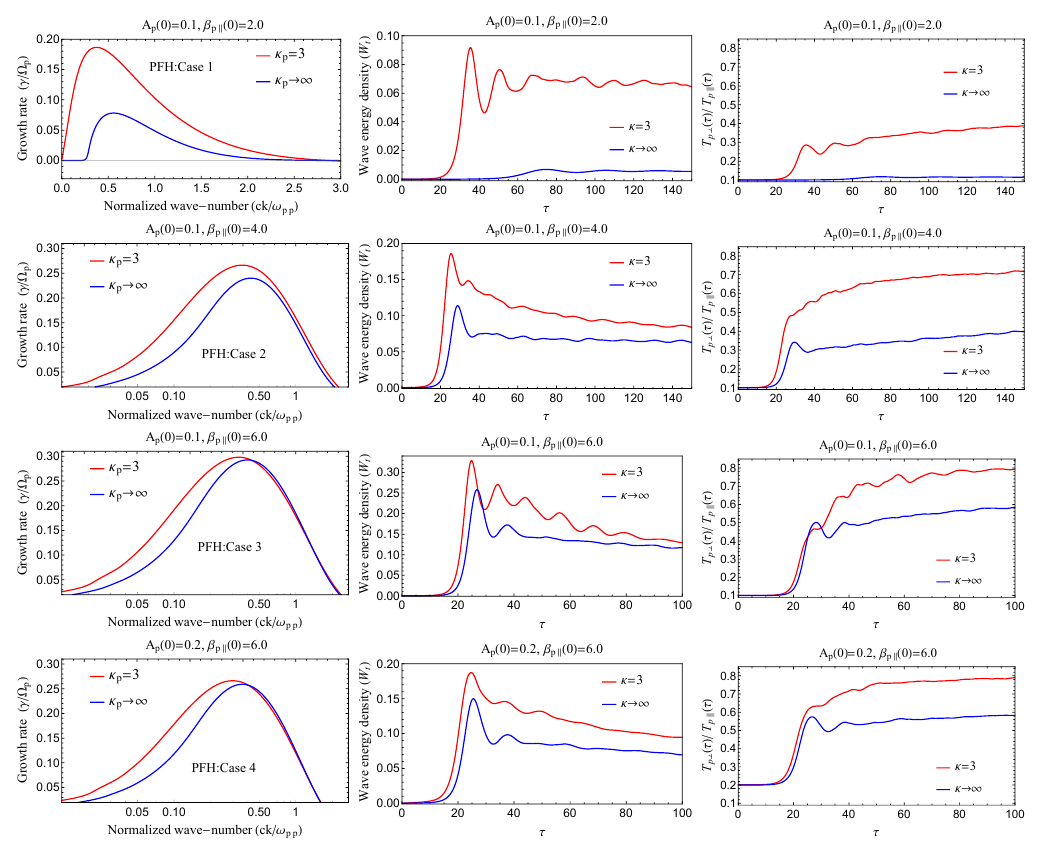}
    \caption{PFH unstable solutions for bi-Kappa protons (red) and the corresponding bi-Maxwellian core (blue) for all three cases in Table~\ref{t2} with initial proton parameters also indicated in the panels. Comparison of growth rates from linear theory (left panels) and temporal evolution from simulations for the wave energy density (middle panels) and the temperature anisotropy (right panels).}
    \label{f4}
\end{figure*}
\begin{figure}[h!]
    \centering
     \includegraphics[width=0.41\textwidth,clip]{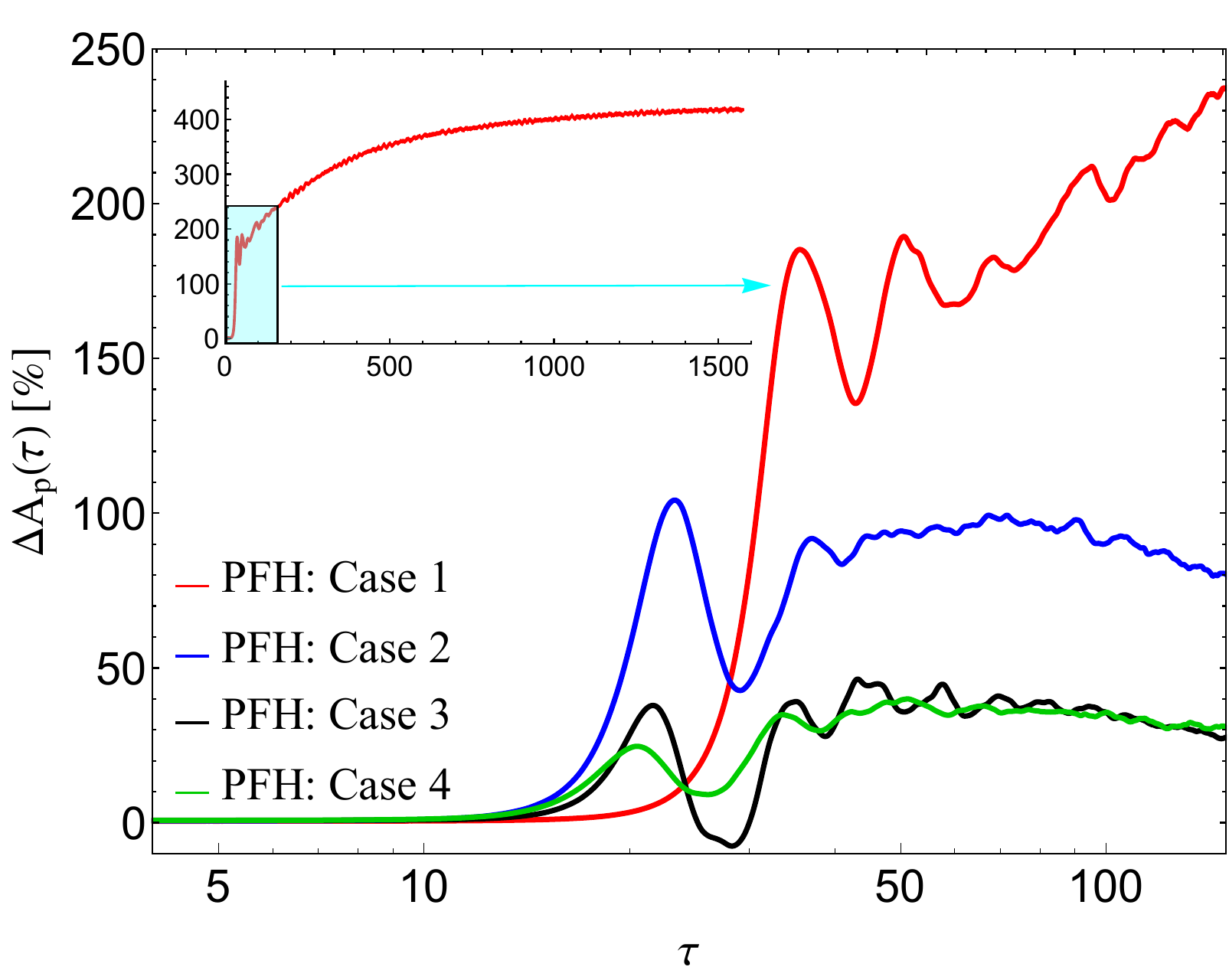}
    \caption{Instantaneous relative difference of the proton anisotropy, $\Delta A_p (\tau)$ (in \%, see definition in the text), induced by the presence of suprathermal protons for all PFH runs in Fig.~\ref{f4}.}\label{f5}
\end{figure}
%
In this section, we discuss the proton firehose (PFH) instability for four cases parameterized in Table~2, in terms of the initial values of the plasma beta parameter, both $\beta_{p\parallel} (0)$ for bi-Maxwellian and $\beta_{p\parallel}^{\kappa = 3} (0) = 2 \beta_{p\parallel} (0)$ for bi-Kappa protons, and the temperature anisotropy $A_p(0) < 1$ (not depending on the distribution model).
Reasonable growth rates of firehose instabilities ($\gamma_m/\Omega_p \geqslant 0.01$) generally require overdense or weakly magnetized plasmas (i.e.,  $\beta_{p\parallel} \geqslant 1$), and if the parallel beta is not very high, for instance, $\beta_{p\parallel} < 10$, PFH instability is triggered by the anomalous resonance of protons \citep{Gary-etal-1998, Astfalk-Jenko-2018}. 
The PFH solutions are displayed in Fig.~\ref{f4} for both bi-Kappa (red) and bi-Maxwellian (blue) models, using the same structure from Fig.~\ref{f1}.
The normalized (linear) growth rates ($\gamma/\Omega_p$) are shown in the left column and exhibit a similar enhancement in the presence of suprathermal protons.
Their stimulation is more significant at low values of the beta parameter, leading to increases of 3 or even 4 times the maximum growth rate if, for example, $\beta_{p\parallel} (0) =2.0$ (case~1, top panel).
However, this effect decreases rapidly with increasing beta parameter, suggesting a lower involvement of the suprathermal component. 
Such that for $\beta_{p\parallel} (0)=6$, the suprathermal influence remains noticeable only at low wave numbers $k \leqslant k_m$ (case~4, bottom panel), where $k_m$ corresponds to the mode with the fastest growth, and peaking growth rate $\gamma_m$.

\begin{table}[t!]
    \caption{Parameterization adopted in the hybrid simulations of the PFH instability, in terms of initial values ($\tau = 0$) for four distinct runs.} \label{t2}
   \centering 
    \begin{tabular}{@{}lcccc}
    \hline
   Parameters $ \backslash $ Runs & 1 & 2 & 3 & 4 \\
    \hline
    $\beta_{p,\parallel}(0)$ & 2.0 & 4.0 & 6.0 & 6 \\
    $\beta_{p,\parallel}^{\kappa=3} (0) $ & 4.0 & 8.0 & 12.0 & 12.0  \\
    $A_p (0)$ & 0.1 & 0.1 & 0.1 & 0.2 \\
    \hline 
    \end{tabular}
\end{table}
\normalsize

Corresponding to the PFH solutions for bi-Maxwellian (blue) and bi-Kappa (red) protons, in the other panels of Fig.~\ref{f4} we compare the time evolution of the wave energy density ($W_t (\tau)$, middle column), and the time relaxation of the temperature anisotropy ($A_p (\tau)$, right column), as obtained from simulations.
The effect of suprathermal protons also translates into faster increases of energy density $W_t (\tau)$ and higher values reached saturation, and even after, in the asymptotic decreasing profiles.
In the presence of suprathermal protons, the anisotropy relaxation is not significantly faster but deeper or closer to isotropy (right panels).
These differences depend on the initial values of the main parameters and are generally conditioned by the anisotropy and the plasma beta parameter.
In case~3, the intense wave fluctuations obtained for bi-Kappa protons with $A_p(0) =0.1$ and $\beta_{p \parallel}(0) = 6.0$ are very efficient in the relaxation process leading to very small deviations from isotropy ($A_p \lesssim 1$), with asymptotic values of only $A_p (\tau \to\infty)=0.8$. 
The results for bi-Maxwellian protons are very similar to those obtained from PIC simulations with a reduced (unrealistic) proton-electron mass ratio \citep{Seough2015}, suggesting a minor dependence on this parameter, as \cite{Seough2015} also claimed.

For the same initial value of the temperature anisotropy, the linear properties and the evolution of the instability are conditioned by the beta parameter and the suprathermal exponent $\kappa$. 
For a low initial plasma beta, for instance, $\beta_{p \parallel} (0)=2$ in case~1, the wave energy density reached in the presence of suprathermal bi-Kappa ($\kappa=3$) protons is much higher (one order of magnitude higher) than for bi-Maxwellian protons. 
As an effect of the enhanced fluctuations, the temperature anisotropy also relaxes faster and deeper. Comparisons between runs~1, 2, and 3 show that these differences introduced by suprathermals in the time evolution of wave fluctuations and temperature anisotropy decrease with increasing the initial plasma beta, for instance, $\beta_{p,\parallel} (0)=4.0$ and $6.0$. 
The explanation can also be found because this instability is specific to magnetized plasmas. 
This means that at low $\beta_{p \parallel}$, where the effects of the magnetic field are more pronounced, the growth rates and, implicitly, the level of fluctuations increase significantly with the increase of $\beta_{p \parallel}$, also reflected in the presence of suprathermal populations. 
Conversely, in weakly magnetized plasmas with high $\beta_{p \parallel}$, the consequences of the same variations of this parameter become less significant.

%
\begin{figure*}[t!]
    \centering
    \includegraphics[width=0.85\textwidth,trim={3.5cm 7.cm 3.5cm 6.5cm},clip]{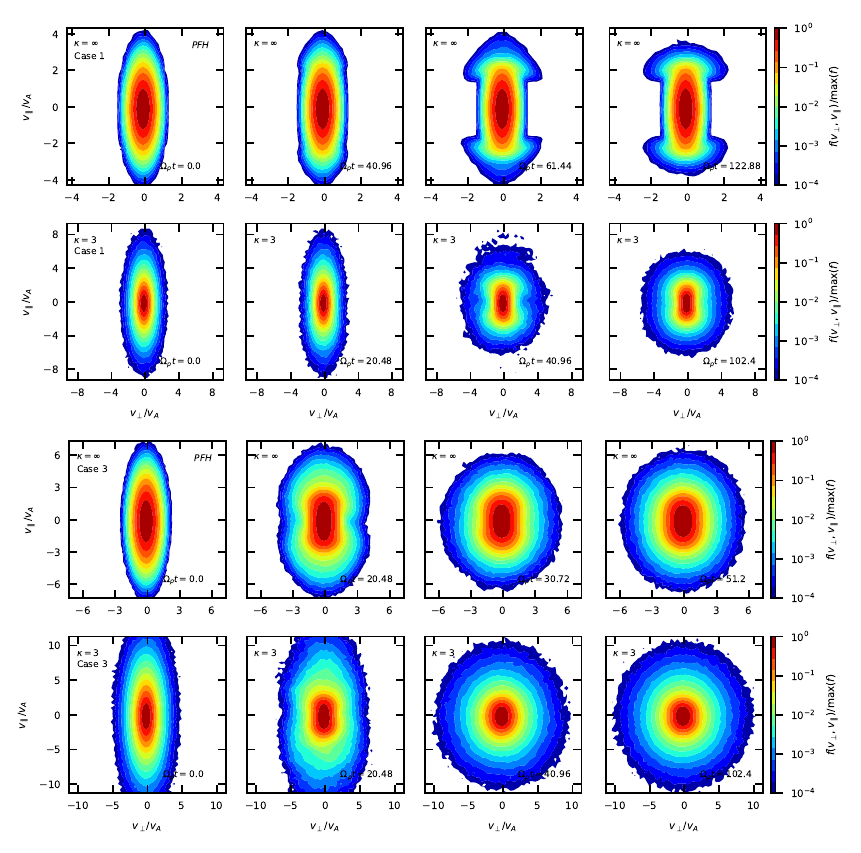}
    \caption{VDs from the simulations of PFH instability for cases~1 and 3 from Figs.~\ref{f4}, computed for four time steps, and for both bi-Maxwellian ($\kappa \to \infty $) and bi-Kappa ($\kappa=3$) distributed protons, as indicated in the panels.}
    \label{f6}
\end{figure*}
%

To quantify the back effect of the enhanced fluctuations on the relaxation of anisotropic protons, we use again the relative difference in the temperature anisotropy $\Delta A_p$, defined in \eqref{eq9}. 
Fig.~\ref{f5} displays the time variation for each run in the right column of Fig.~\ref{f4}.
This also enables us to examine the impact of the chosen initial plasma parameters, including the indirect influence of suprathermal protons on the relaxation of temperature anisotropy. 
The instantaneous relative differences $\Delta A_p~(\%$) increase as time evolves, reaching the first peaks at the early stages, specifically at $\tau \in [20-35]$, and continuing to fluctuate until the end of the simulations. 
The first peaks indicate that for suprathermal protons, the relaxation of temperature anisotropy starts faster than that for less energetic protons. 
For run~1 (red line) with a low initial plasma beta $\beta_{p \parallel} (0)=2$, the first peak $\Delta A_p \approx 185 \%$ appears at $\tau=35$, before $\Delta A_p$ grows and eventually reaches $235 \%$ at $\tau=150$. In other words, in the presence of suprathermal protons, the relaxation of the temperature anisotropy is $185\%$ and $235\%$ deeper at, respectively, $\tau=35$ and $\tau=150$. 
The subfigure (left corner) in Fig.~\ref{f5} shows the saturation from longer runs of $\tau_{max}=1570$, at about $\Delta A_p(\tau_{max})\approx 416\%$, and $A_p(\tau_{max})\approx 0.6$, concurrent with instability saturation (not shown here).
With increasing the initial $\beta_{p \parallel}(0)$, the contrast introduced by the suprathermal population is reduced for the abovementioned reasons. 
For instance, in run~2 with $\beta_{p \parallel}(0)=4$, $\Delta A_p$ displays its first peak of value $103\%$ at $\tau=24$, before eventually reaching the value $80\%$ at $\tau=150$. 
Furthermore, for $\beta_{p \parallel}(0)=6$ in run~3 the first peak of $\Delta A_p$ reduces to $38\%$ at $\tau=22$, before reaching a lower value of $29\%$ at $\tau=150$. 
Suprathermal populations have their highest potential at lower plasma beta. 
The initial plasma parameters in run~4 are similar to those of run~3 but with lower anisotropy (two times lower). 
In this case, the time evolution of $\Delta A_p$ is similar to that in run~3, especially at later stages (e.g., at $\tau>30$), reaching a value of $28\%$ at $\tau=150$.

Fig.~\ref{f6} displays snapshots of proton VDs from the simulations, which enable us to trace, as time evolves, the deformations undergone by the distributions by comparison to their initial ($\tau=0$) shape in response to the exponential growth of PFH fluctuations.
We follow the same structure as in Fig.~\ref{f3}, and the upper two rows of panels contrast the bi-Maxwellian ($\kappa\rightarrow \infty$) and bi-Kappa ($\kappa=3$) protons from run~1 in Fig.~\ref{f2} (i.e., for $\beta_{p \parallel}=2$ and $6$). 
In these cases, the initial shape of the bi-Maxwellian or bi-Kappa distributions may not be preserved during the relaxation of temperature anisotropy.
The departure from the initial shape of VD is a dumbbell-like deformation in the parallel direction, with the excess of kinetic energy ($T_\parallel > T_\perp$). These deformations are more prominent for bi-Maxwellian distributed protons, and those low-beta protons, see the upper panels in Fig.~\ref{f6}. Dumbbell-like deformations were also reported by \cite{Matteini-etal-2006} and \cite{Seough2015} from the simulations of the PFH instability of similar low-beta bi-Maxwellian protons.
Even in bi-Kappa distributions, the dumbbell-like deformation can persist longer for low-energy protons in the core of the VD, similar to the low-beta population. 
A plausible explanation invoked by \cite{Matteini-etal-2006} is the distinctive resonance of the growing fluctuations with the proton populations, shaping only part of the VD.
The diffusion of more energetic particles (and implicitly less magnetized particles) is much faster and has fewer significant deformations. 
This is reflected both in the distributions of high-beta protons, compare, for instance, runs~1 and 3 in Fig.~\ref{f6}, but also in the (initial) bi-Kappa distributions contrasting with bi-Maxwellian ($\kappa \to \infty$).
In this sense, the case of bi-Kappa protons with $\beta_{p \parallel} = 6$ (lower panels) show the most pronounced relaxation as time evolves.

\subsection{Simulations vs. linear theory}\label{LVS}
%
\begin{center}
\begin{figure}
    \centering
    \includegraphics[width=0.48\textwidth,trim={3.5cm 5.cm 3.5cm 3.cm},clip]{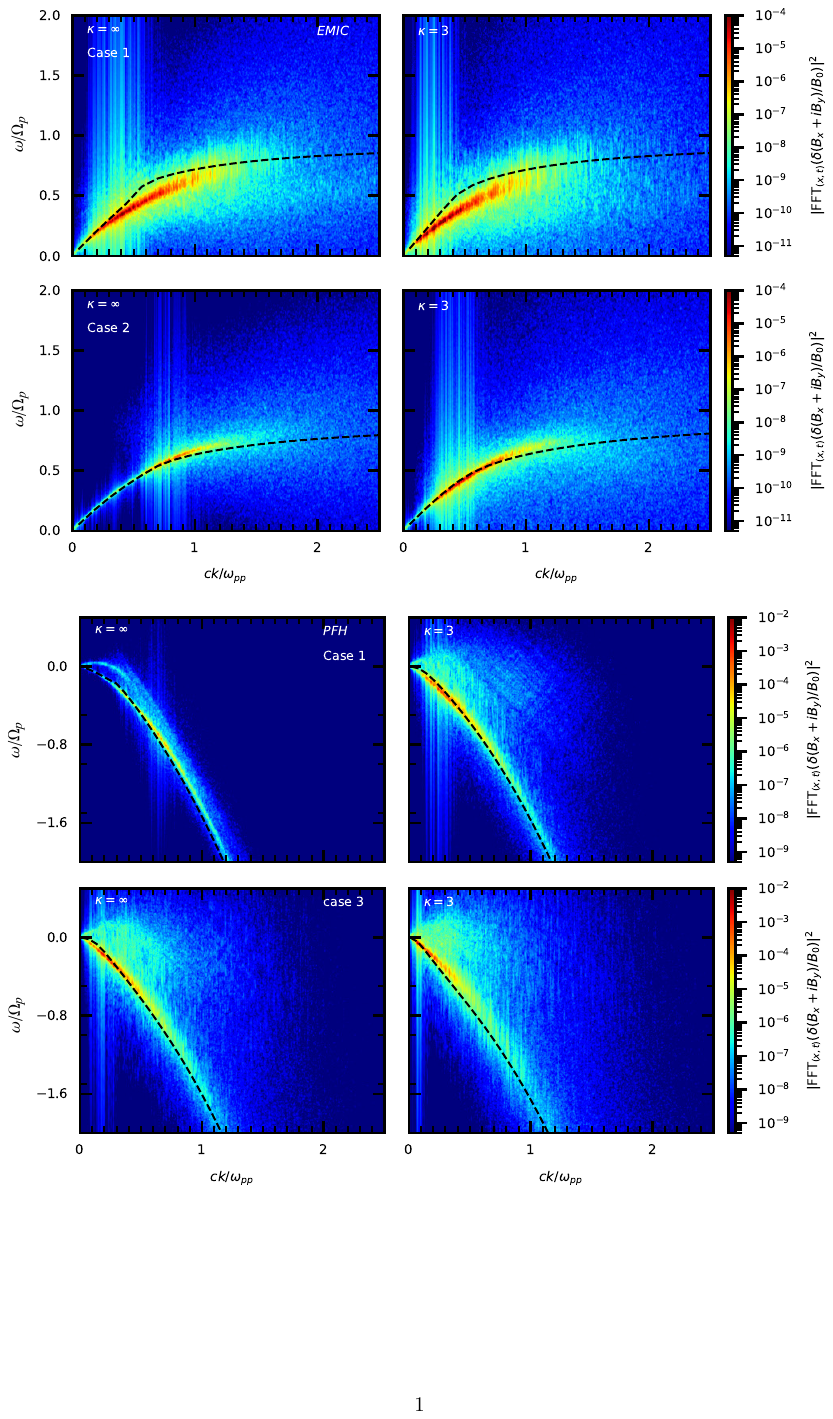}
    \caption{Magnetic fluctuations from simulations, for EMIC cases~1 and 2 (upper panels) and PFH instability cases~1 and 3 (lower panels), resemble linear dispersion (dashed lines) in the $\omega - k$ space and are enhanced in the presence of suprathermals (right panels). The colour scale indicates the normalized magnetic wave energy density.} \label{f7}
\end{figure}
\end{center}
%

Here, we use two methods to test the numerical simulation results against the linear theory predictions for both instabilities.
On the one hand, Fig.~\ref{f7} presents examples of magnetic fluctuation spectra obtained from simulations in the $\omega-k$ space, by computing the fast Fourier transforms in space and time as $|\text{FFT}_{(x,t)}\{\delta(B_x+iB_y)/B_0\}|^2$, for EMIC instability cases~1 and 2 (upper panels), and for PFH instability cases~1 and 3 (lower panels). 
The results for bi-Maxwellian protons are in the left panels, and for bi-Kappa in the right panels.
In this way, the left-handed polarized fluctuations of EMIC instability are shown in the first quadrant [$\omega>0$, $k>0$]. In contrast, the right-handed ones of PFH instability appear in the fourth quadrant [$\omega<0$, $k>0$].
The suprathermal protons enhance peak intensities associated with the triggered instabilities; see the right panels. 
Moreover, these intensities align to the dispersion curves $\omega_r(k)$ derived from linear theory (over-plotted dashed lines), showing satisfactory agreement for all cases.
%
\begin{center}
\begin{figure} [t!]
    \centering
    \includegraphics[width=0.48\textwidth,trim={2cm 14cm 1.5cm 2.2cm},clip]{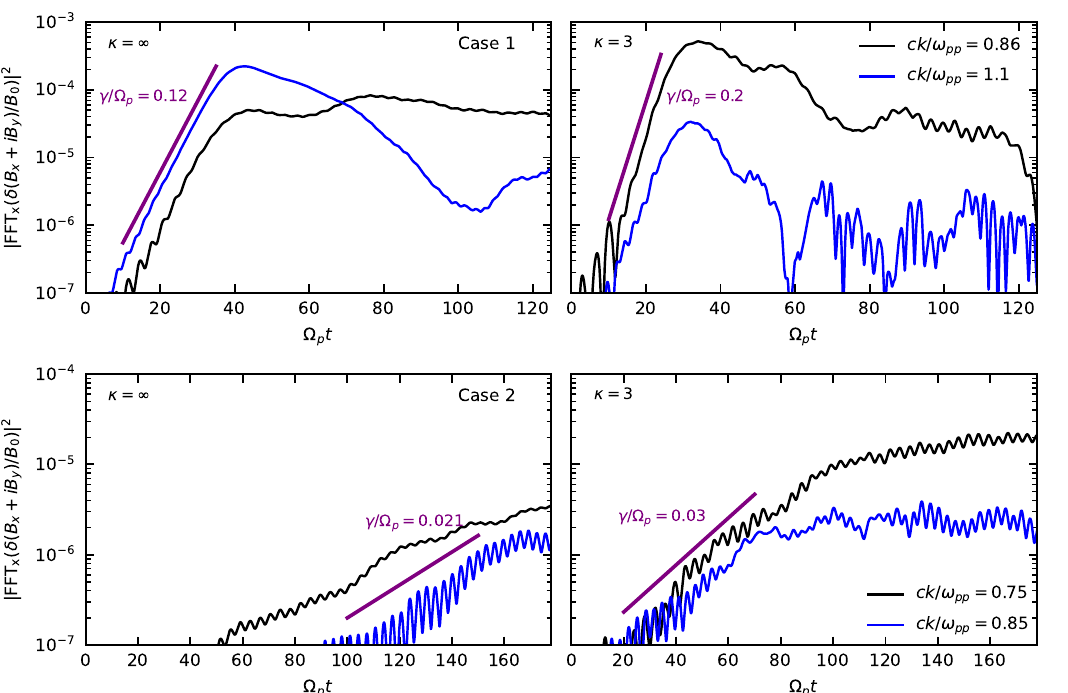}
    \caption{Time variation of magnetic fluctuation spectra for two EMIC modes corresponding to distinct wave numbers (see the legend), in cases~1 and 2. Computed as the spatial Fourier transform, $|\text{FFT}_x\{\delta(B_x+~iB_y)/B_0\}|^2$.}
    \label{f8}
\end{figure}
\end{center}
\begin{center} 
\begin{figure}[h!]
    \centering
    \includegraphics[width=0.48\textwidth,trim={2cm 14cm 1.5cm 2.2cm},clip]{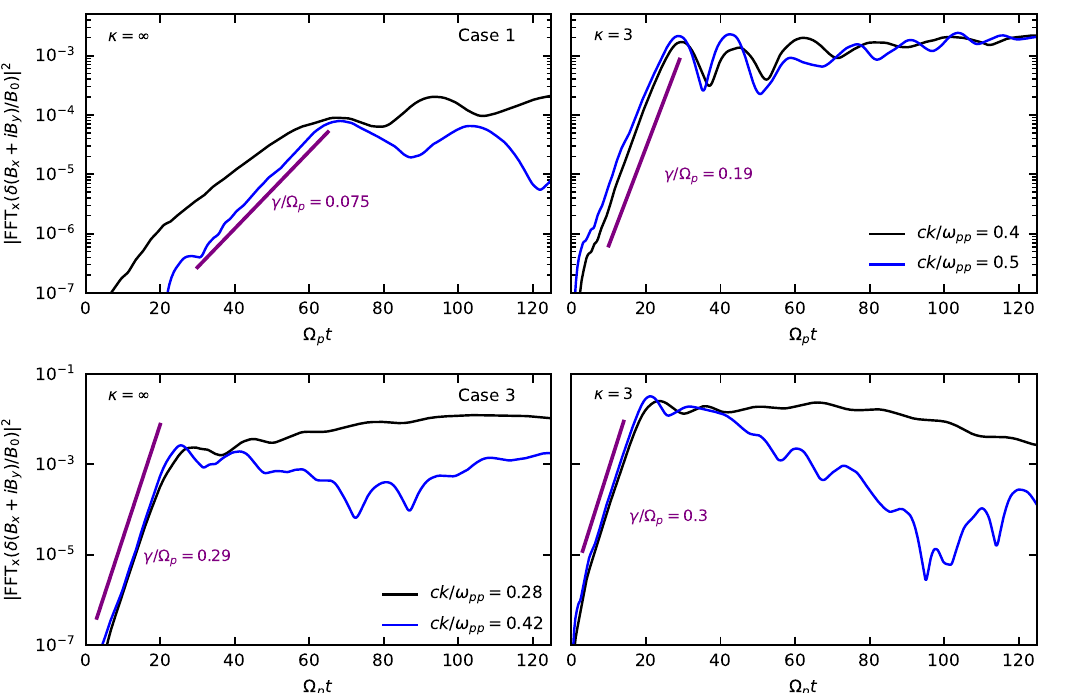}
    \caption{Time variation of magnetic fluctuation spectra for two PFH modes corresponding to distinct wave numbers (see the legend), in cases~1 and 3.}
    \label{f9}
\end{figure}
\end{center}
%

On the other hand, we can also examine the time variation of magnetic fluctuation spectra, computed as the spatial Fourier transforms $|\text{FFT}_x\{\delta(B_x+iB_y)/B_0\}|^2$ for two distinct modes of EMIC instability, in, for example, cases 1 and 2 shown in Fig.~\ref{f8}, and for other two modes of PFH instability in cases 1 and 3 displayed in Fig.~\ref{f9}. 
The wave numbers of these modes, distinguished by blue and black lines, are indicated in the legends. 
The purple lines have slopes given by the growth rates of the same wave modes obtained from linear theory, that is, $\propto \exp(2\gamma t)$, with growth rate $\gamma/\Omega_p$ given in Figs.~\ref{f1} and \ref{f4} (also indicated along the purple lines).
For both instabilities in Figs.~\ref{f8} and \ref{f9}, as well as for both bi-Maxwellian and bi-Kappa distributed protons, the initial growth of the magnetic fluctuation is sufficiently well predicted by these growth rates from linear theory.
%
\section{Conclusions}\label{sec5}

This paper reports the first comprehensive results from simulations of EM instabilities triggered by protons with anisotropic temperatures modelled by bi-Kappa distributions. 
Suprathermal plasma populations, electrons and protons enhance the high-energy (bi-)Kappa tails of the velocity or energy distributions measured in situ in space plasmas. Still, their effect is often ignored, limiting the analysis to the quasi-thermal population in the low-energy (bi-)Maxwellian core. 
Recent comparative analyses, summarized for instance in \cite{Shaaban-etal-2021-book} and \cite{Lazar-etal-2022}, show that suprathermal populations stimulate instabilities generated by temperature anisotropy, such as the EM ion cyclotron (EMIC) and proton firehose (PFH) instabilities. 
This appears to be a systematic effect, which has already been demonstrated in theory and numerical simulations for electrons.

Here, we have added culminating results to this series of investigations, this time using hybrid simulations that confirm that suprathermal protons indeed stimulate both EMIC and PFH instabilities.
The effects are also consistent from a theoretical point of view because the suprathermal protons contribute to an increase in the kinetic energy density and, therefore, to an increase in the beta parameter.
For both instabilities, the resulting wave fluctuations can reach significantly enhanced levels of wave energy density at saturation, and the effects of suprathermal protons are most striking when the proton beta parameter is low. 
Both the EMIC and PFH instabilities are typical of magnetized plasmas, meaning that in a low proton beta regime (i.e., $\beta_p \lesssim 1$ for EMIC), these instabilities respond with larger variations in growth rate and wave energy levels than in the regimes of weakly magnetized plasmas with a high plasma beta (i.e., $\beta_p > 1$ for EMIC).
The enhanced levels of fluctuations obtained in the presence of suprathermal protons, in turn, cause faster and/or deeper relaxation of the temperature anisotropy. 
Anisotropic protons with bi-Kappa distributions generally show such a fast and deeper relaxation, and it contrasts markedly with the relaxation of the bi-Maxwellian core when the involvement of the suprathermal component is neglected.
The PFH instabilities tend to distort the bi-Kappa distributions of protons with the same dumbbell-like deformations previously shown for the bi-Maxwellian distributions. 
However, these deformations are less prominent for bi-Kappa protons, and the relaxation is more uniform.

At first glance, 1D simulations seem to have a major limitation, but parallel propagating instabilities are the fastest relaxation channels for the chosen regimes.
Thus, for protons with anisotropy $A > 1$ and $\beta_{p \parallel} < 5$, the parallel EMIC develops with higher growth rates than the oblique modes, including the mirror mode \citep{Gary-etal-1976, Yoon-Seough-2012, Yoon-etal-2023ApJ, Shaaban-etal-2018-mirror}.
Although limited to only one set of parameters, a recent result confirms higher EMIC growth rates also in the presence of suprathermal protons \citep{Lopez-etal-2023}. 
Instead, we presented extensive results from analysing different regimes varying the relevant parameters. 
Moreover, the parallel PFH instability triggered by $A < 1$ benefits from regimes of extended relevance when their growth rate exceeds that of the oblique modes \citep{Micera2020, Lopez_2022}. 
The parametric regimes in Table~2 are all favorable to parallel PFH instability, see Figure 3 (left panel) from \cite{Lopez_2022}. 
Note that in our case~1, the presence of suprathermal protons shifts from the dominance of oblique PFH to that of parallel PFH instability due to the increase in plasma beta.
Restricting our analysis to parallel-propagating instabilities is supported by theory and solar wind observations of proton-scale fluctuations \citep{Woodham_2019}, as explained in the introduction.

As we have examined here, these conclusions are valid for multiple regimes with varied parameters, conferring extended relevance not only in space plasmas such as the solar wind and planetary magnetospheres but very likely in other similar astrophysical contexts as well.
Moreover, such numerical analyses offer realistic perspectives for modelling the dynamics of non-equilibrium plasmas, in which the suprathermal populations are an essential component intimately connected with the kinetic wave fluctuations.
At the same time, these results should motivate further, more detailed investigations of the complete spectra covering the 2D space of the wave vectors to clarify the possible involvement of the other oblique modes under the influence of the suprathermal populations.
Future studies could also use improved hybrid simulations, with, for example, particle test methods \citep{Trotta-etal-2019}, to elucidate, for instance, the potential role of the (enhanced) proton instabilities in the acceleration of electrons from interplanetary shocks, where suprathermal protons are ubiquitous \citep{Lario2019, Yang-etal-2023}.

\section*{Acknowledgements}

The authors acknowledge support from the Ruhr-University Bochum, the Katholieke Universiteit Leuven, and Qatar University. These results were also obtained in the framework of the projects C14/19/089 (C1 project Internal Funds KU Leuven), G002523N (FWO-Vlaanderen), 4000145223 (SIDC Data Exploitation (SIDEX2), ESA Prodex), Belspo project B2/191/P1/SWiM. Powered@NLHPC: This research was partially supported by the supercomputing infrastructure of the NLHPC (ECM-02).
We thank the anonymous reviewer for constructive criticism and suggestions. 
\bibliographystyle{aa}

\bibliography{papers}

\end{document}